# Enhanced third harmonic generation from the epsilon-near-zero modes of ultrathin films

Ting S. Luk[1,2,*], Domenico de Ceglia[3,*], Sheng Liu[1,2], Gordon A. Keeler[1], Rohit P. Prasankumar[1,4], Maria A. Vincenti[3], Michael Scalora[5], Michael B. Sinclair[1], and Salvatore Campione[1,2]

[1] Sandia National Laboratories, P.O. Box 5800, Albuquerque, NM 87185, USA

[2] Center for Integrated Nanotechnologies (CINT), Sandia National Laboratories, P.O. Box 5800, Albuquerque, NM 87185, USA

[3] National Research Council – AMRDEC, Charles M. Bowden Research Laboratory, Redstone Arsenal, AL 35898 USA

[4] Center for Integrated Nanotechnologies (CINT-LANL), Los Alamos Laboratories, Los Alamos, NM 87545 USA

[5] Charles M. Bowden Research Laboratory, AMRDEC, US Army RDECOM, Redstone Arsenal, AL 35898 USA

[*] These authors contributed equally to this work and are joint first authors

Correspondence to T.S.L. (tsluk@sandia.gov)

ABSTRACT

We experimentally demonstrate efficient third harmonic generation from an indium tin oxide (ITO) nanofilm ($\lambda/42$ thick) on a glass substrate for a pump wavelength of 1.4 μm. A conversion efficiency of $3.3 \times 10^{-6}$ is achieved by exploiting the field enhancement properties of the epsilon-near-zero (ENZ) mode with an enhancement factor of 200. This nanoscale frequency conversion method is applicable to other plasmonic materials and reststrahlen materials in proximity of the longitudinal optical phonon frequencies.

Third harmonic generation is a commonly used nonlinear optical process that triples the input photon energy. Large conversion efficiency in traditional nonlinear optical devices requires large nonlinearities, low material absorption, and phase-matching techniques that increase the interaction length to the millimeter-to-centimeter range. Phase matching is irrelevant at the nanoscale, and new strategies must be developed to boost the performance of sub-wavelength nonlinear optical devices which are expected to play an important role in optoelectronics and optical information processing. Several approaches involving high-Q photonic modes have been proposed, including ring cavity modes, guided mode resonances, photonic crystal band edges, and defect states of periodic structures[1]. Recently, metallic and more generally plasmonic nanostructures have received considerable attention[2, 3]. While the Q-factors of plasmonic resonators are usually smaller than those achieved with all-dielectric photonic devices, larger field enhancements are possible since plasmonic modal volumes can be deeply sub-wavelength. However, harmonic generation using the sub-wavelength structures reported thus far relies on field enhancements associated with localized surface plasmon resonances or collective resonances and requires exquisite fabrication techniques[1, 4-7].

Frequency-mixing from interfaces and thin nonlinear films has also been intensely investigated since the early days of nonlinear optics[8, 9], with several studies exploiting the field enhancement associated with the excitation of short[10] and long[11] range surface plasmon polaritons[12]. More recently, intriguing light-matter interactions occurring in natural or artificial epsilon-near-zero (ENZ) materials have come under scrutiny. In this paper we present a method to enhance third harmonic (TH) generation using the ENZ polariton[13] supported by deeply sub-wavelength, un-patterned films. The material utilized in this work is indium tin oxide (ITO), a common transparent conductive oxide, but the results can be generally extended to other materials such as reststrahlen materials that exhibit ENZ behavior and nonlinear properties.

In natural media, epsilon near zero behavior occurs at the plasma frequency — the frequency at which the real part of a material's dielectric constant crosses zero. Plasma frequencies in the ultraviolet and visible ranges are typical for metals, while heavily-doped semiconductors or oxides such as ITO show zero-crossing frequencies in the near-infrared. Moreover, ENZ may be synthesized at virtually any frequency with properly designed metamaterials, using sub-wavelength arrangements of plasmonic resonators or using guided modes and operating near the cutoff frequency. Many optical effects and potential applications arising from ENZ behavior have been proposed or demonstrated including optical tunneling[14-16], phase patterning[17], directional emission[18], perfect absorption[19, 20], dielectric sensing[21], guided index lensing[22], enhanced emission[23-25], optical cloaking[26], strong coupling phenomena[27-29], optical modulation[30, 31], thermo-photovoltaics[32], and enhanced optical nonlinearities[33-35].

Bulk plasmon modes (which occur at the ENZ frequency) in infinite homogenous media are longitudinal in nature and cannot interact with transverse electromagnetic fields. However, in ultrathin plasmonic materials an external plane wave may couple to the volume plasmon[36]. When this coupling is achieved to the right of the light line, the volume plasmon polariton mode is sometimes referred to as the epsilon-near-zero mode[13]. Under the right conditions, an external plane wave can be coupled to the ENZ mode and completely absorbed in a deeply sub-wavelength film[20]. For thin-film configurations that exhibit low reflectivity near the ENZ frequency, the continuity of the normal component of the electric displacement necessarily requires the existence of a large electric field immediately inside the film[37]. However, such low reflectivity can only be achieved through excitation of a thin film resonance, which in the present case is the ENZ mode. This field enhancement can, in turn, lead to substantial increases in nonlinear optical processes such as harmonic generation which depend superlinearly on the field amplitude at the fundamental frequency. This concept has been proposed theoretically as a simple and effective way to enhance second and third harmonic generation in ENZ slabs illuminated by p-polarized light at oblique incidence in a symmetric substrate/superstrate configuration[33]. In this work, we propose and experimentally demonstrate an approach to strengthen the coupling to the ENZ mode and further increase the nonlinear response. In particular, we show that excitation of the ENZ mode under total reflection conditions (i.e. above the glass/air critical angle) enhances the reflected third harmonic generation by two orders of magnitude due to ENZ field enhancement. Note that, although the structure in ref. 13 is different from the one employed in this work (metal substrate in Ref. 13 and glass in the present work), many of the important properties of the ENZ modes are similar for the two configurations.

The ENZ sample utilized in this study was purchased from Delta Technologies and consists of a 33 nm thick ITO layer deposited on alumino-silicate glass (Fig. 1a). In the frequency ranges investigated, ITO has a moderately large third order susceptibility which is comparable to that of Si or GaAs. The origin of the large third order nonlinear susceptibility of ITO[38-40] is due to delocalized electrons with large mobility, analogous to pi-electrons in polymers[39]. Schematics of the excitation geometry and optical setup are shown in Fig. 1a and Fig. 1b. The dielectric permittivity of the ITO film was determined from a model fit to measured ellipsometry data (see Fig. 1c), and the ENZ wavelength is determined to be 1.385 μm (7220 cm$^{-1}$). For efficient TH generation, we used the Kretschmann excitation scheme in order to optimize the coupling to the ENZ mode. The signature of efficient coupling to the ENZ mode is the near-perfect absorption feature of incident p-polarized light at the specific angle of incidence of 43.7 degrees. The measured absorption (Fig. 1d) agrees well with the transfer matrix calculation based on the fitted dielectric function.

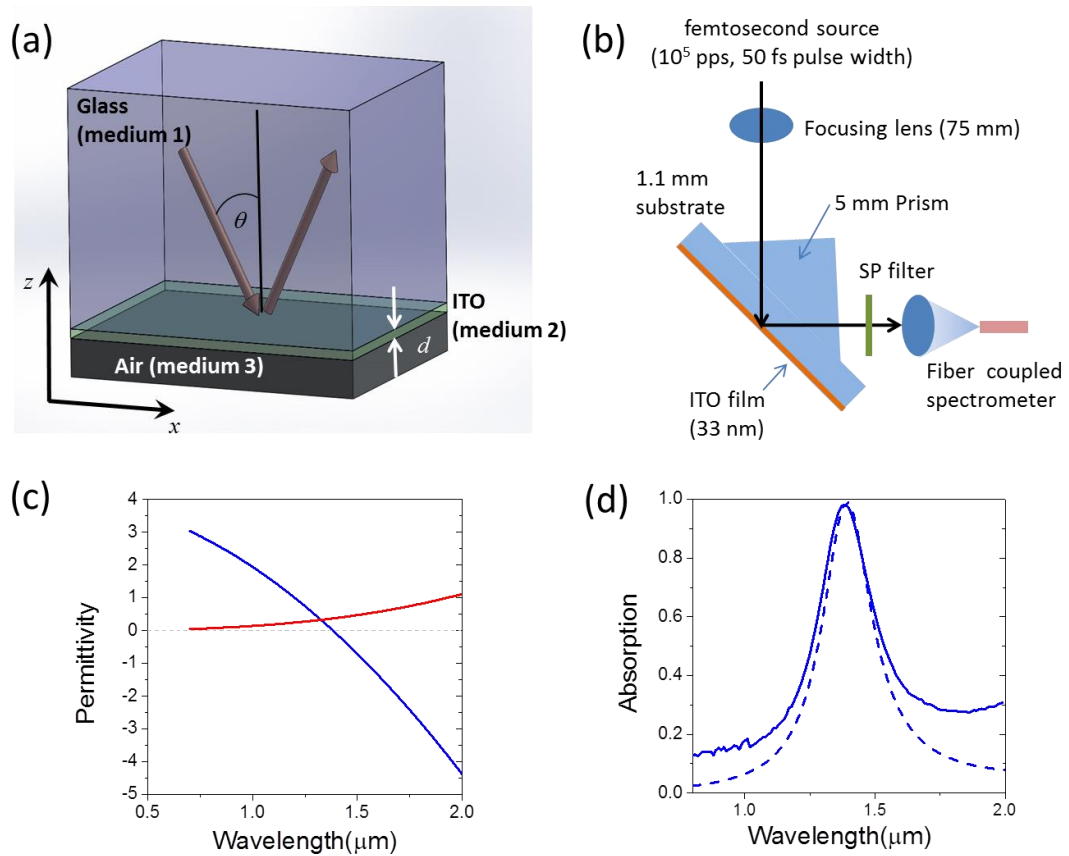

Fig. 1. (a) Schematic picture of the Kretschmann excitation geometry. (b) Essential components of the optical setup for the third harmonic measurement. The short pass (SP) filter used is a 2 mm thick Schott glass BG40 filter. (c) Real (blue) and imaginary (red) parts of the permittivity of the ITO sample versus wavelength. (d) Measured (solid blue) and calculated (dashed blue) p-polarized absorption profile at 43.7 degrees versus wavelength.

The fundamental pump beam is derived from a femtosecond optical parametric amplifier pumped by a Ti:Sapphire regenerative amplifier. The pump wavelength can be tuned in the vicinity of the ENZ wavelength of ITO. This source delivers an average power of about 6-10 mW (depending on the wavelength) with a pulse repetition rate of 100 KHz and a typical pulse width of 50 fs FWHM. The beam was focused with a 75 mm focal length lens to a near Gaussian spot, with full width at $1/e^2$ of about 80 µm as measured using an up-conversion CCD camera. We use a 5 mm prism (matched to the glass substrate) in the Kretschmann geometry to excite the ENZ mode near the glass-to-air critical angle. Due to in-plane momentum conservation and low dispersion of the prism, the reflected third harmonic wave emerges nearly collinearly with the specularly reflected fundamental wave. The harmonic light is visible to a dark-adapted naked eye when projected on a white card. After the fundamental wave is rejected using a 2 mm thick BG40 filter, the third harmonic radiation is collected by a lens and focused onto a fiber-coupled imaging spectrometer equipped with a liquid nitrogen cooled CCD camera. The inset of Figure 2a shows the TH spectrum for the case of a 1.4 µm fundamental wavelength. As expected for a third harmonic process, the spectrum peaks at ~470 nm. Also as expected for third harmonic generation, the intensity of the third harmonic output has a pump intensity dependence of $I_{FF}^{3.0\pm0.15}$, as shown in Fig. 2a. To verify that the harmonic yield depends on the coupling to the ENZ mode, we varied the incident wavelength and angle, and in all cases the third harmonic yield declines when the excitation condition deviates from the optimal coupling condition (see Fig. 2b). A mere shift of the pump wavelength by 100 nm from the optimum 1.4 µm to 1.3 µm reduces the TH emission by nearly a factor of 2. At the maximum average pump power of 6 mW (pulse peak intensity of $2 \times 10^{10}$ W/cm$^2$), we obtained 20 nW of third harmonic, which implies a conversion efficiency of $3.3 \times 10^{-6}$, an impressive value given the sub-wavelength dimensions of the ITO layer. The third harmonic efficiency was also measured in the non-Kretschmann geometry, with incidence from air onto the ITO films, and was found to be ~200 times smaller than the TH from the Kretschmann geometry (see Fig. 2c). Finally, the measurements were repeated without the ITO sample in place and we found that the TH generation from the prism glass only was more than four orders of magnitude smaller than the ITO contribution.

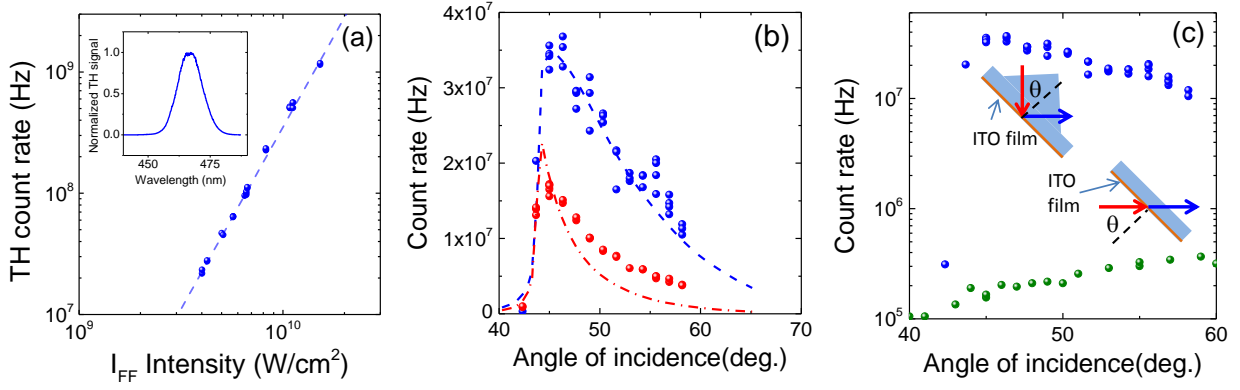

Fig. 2. (a) $I_{FF}$ intensity dependence of third harmonic yield for a 1.4 μm (7246 cm$^{-1}$) incidence wavelength. The inset shows the third harmonic spectrum. (b) The angular dependence of third harmonic yield as a function of incident angle for incidence wavelengths of 1.3 μm (red dots) and 1.4 μm (blue dots) at $I_{FF}$ intensity of 1.2x10$^{10}$ W/cm$^2$. The dashed lines are full-wave simulation results for corresponding excitation conditions. Due to the uncertainty of the absolute experimental angle, the theoretical result was shifted by 2 degrees. (c) TH yield versus angle of incidence of 1.4 μm pump wave for the Kretschmann (blue dots) and non-Kretschmann (green dots) geometries. The excitation geometries are also shown as insets.

From a theoretical perspective, the waves radiated at the third harmonic frequency can be obtained by solving the inhomogeneous Helmholtz equation using the nonlinear polarization as the source term:

$$\nabla \times \nabla \times \mathbf{E}_{TH} - \omega_{TH}^2/c^2 \boldsymbol{\varepsilon}_{TH} \cdot \mathbf{E}_{TH} = \omega_{TH}^2 \mu_0 \mathbf{P}_{TH} \qquad (1)$$

where $\mathbf{E}_{TH}$ is the TH electric field, $\boldsymbol{\varepsilon}_{TH}$ is the relative permittivity tensor at the TH frequency, $\omega_{TH}$ is the TH angular frequency and $\mathbf{P}_{TH}$ is the TH nonlinear polarization density, i.e., the source of the TH signal. The solution of this inhomogeneous equation can be expressed as the superposition of a TH free wave that is the solution of the homogeneous wave equation (assuming $\mathbf{P}_{TH} = \mathbf{0}$ in the Helmholtz equation) and travels at the TH phase and group velocity, and TH bound waves that are particular solutions related to the presence of the inhomogeneous term $\omega_{TH}^2 \mu_0 \mathbf{P}_{TH}$ and locked to the pump field[8, 41].

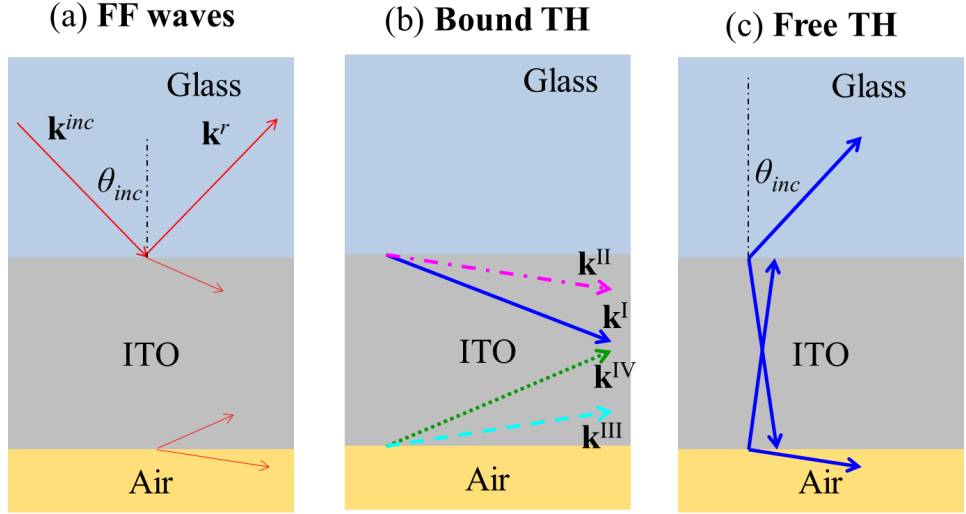

Fig. 3 (a) Fundamental frequency waves, (b) bound TH waves, and (c) free TH waves in the input medium, ITO film, and output medium. In (b), the arrows labeled with roman numbers correspond to the bound, inhomogeneous TH sources in Eq. (2). In (c), the blue solid arrows represent the free, homogeneous TH components.

The source term appearing in the Helmholtz equation (1) stems from the nonlinear mixing of the forward and backward pump waves in the ENZ film which generates four polarization waves at the third harmonic frequency which are *locked* to the fundamental electric fields. Thus, the nonlinear polarization density at the TH frequency is given by:

$$\mathbf{P}_{TH} = e^{(i3k_x^{inc}x - i3\omega_{FF}t)}$$
$$\varepsilon_0 \, [\chi^{(3)} \cdot \mathbf{E}_+\mathbf{E}_+\mathbf{E}_+ e^{i3k_z z} +$$
$$3\chi^{(3)} \cdot \mathbf{E}_+\mathbf{E}_+\mathbf{E}_- e^{ik_z z} + \quad (2)$$
$$3\chi^{(3)} \cdot \mathbf{E}_+\mathbf{E}_-\mathbf{E}_- e^{-ik_z z} +$$
$$\chi^{(3)} \cdot \mathbf{E}_-\mathbf{E}_-\mathbf{E}_- e^{-i3k_z z}]$$

where $k_x^{inc}$ is the transverse component of the incident fundamental-frequency (FF) wave-vector, $\omega_{FF}$ is the FF angular frequency, $k_z$ is the longitudinal (z-direction) component of the FF wave-vector in the ITO film and $\mathbf{E}_{+/-}$ are the complex amplitudes of the forward (+) and backward (−) FF waves (see Fig. 3 for a pictorial view of these waves). In the undepleted pump approximation (i.e. the TH generation does not cause pump depletion), $\mathbf{E}_{+/-}$ may be found using the linear transfer matrix technique at the FF. In Eq. (2) we can identify a purely forward bound wave with wavevector $\mathbf{k}^I = 3k_x^{inc}\hat{\mathbf{x}} + 3k_z\hat{\mathbf{z}}$ and amplitude dependent only on the forward component of the pump wave and a purely backward wave with wavevector $\mathbf{k}^{IV} = 3k_x^{inc}\hat{\mathbf{x}} - 3k_z\hat{\mathbf{z}}$ with amplitude dependent on the backward component of the pump signal. The other two bound waves originate from the mixing of the forward and backward pump waves. Hence their wavevectors are $\mathbf{k}^{II,III} = 3k_x^{inc}\hat{\mathbf{x}} \pm k_z\hat{\mathbf{z}}$. The polarization waves then radiate two homogeneous or *free* third harmonic waves that propagate in the forward and backward directions following the usual Snell's refraction law in the ITO film. Interestingly, we find that the four inhomogeneous waves in Eq. (2) are equally important for the nonlinear interaction when the ENZ mode is excited, and give similar contributions to the overall TH generation efficiency.

Solution of the inhomogeneous Helmholtz equation (1) for the fields radiated at the harmonic frequency[8] computed via full-wave simulations are shown in Fig. 2b as dashed lines and provide excellent agreement with the experimental results. In these simulations we assumed an isotropic nonlinear tensor with only three nonzero terms: $\chi_{xxxx} = \chi_{yyyy} = \chi_{zzzz} = \chi$, and achieved best agreement with the experimental results when $\chi = 3 \cdot 10^{-21}$ m$^2$/V$^2$, which is in good agreement with previous estimations[39]. In Fig. 4 we show maps of theoretical TH conversion efficiencies as a function of frequency and angle of incidence obtained via full-wave simulations. The efficiency is calculated as $\eta^{T,R} = P_{TH}^{T,R} / P_{FF}$, where $P_{TH}^{T,R}$ is either the TH power transmitted to the air side or reflected back to the prism, and $P_{FF}$ is the input FF power. Both efficiency maps display peaks near the ENZ crossing point of ITO (~1.385 μm or 7220 cm$^{-1}$), where pump absorption and field enhancement are maximized. The peaks also have angular selectivity, showing a strong maximum close to the glass/air critical angle (~41.8 degrees, indicated as white dashed lines in the theoretical maps). The transmitted TH peak just below the critical angle is more than an order of magnitude smaller than the reflected TH peak a few degrees above the critical angle. Note that these predictions are valid when the TH process does not deplete the pump and provided nonlinear saturation and self-phase modulation effects are not significant — approximations that are amply justified in the case under investigation.

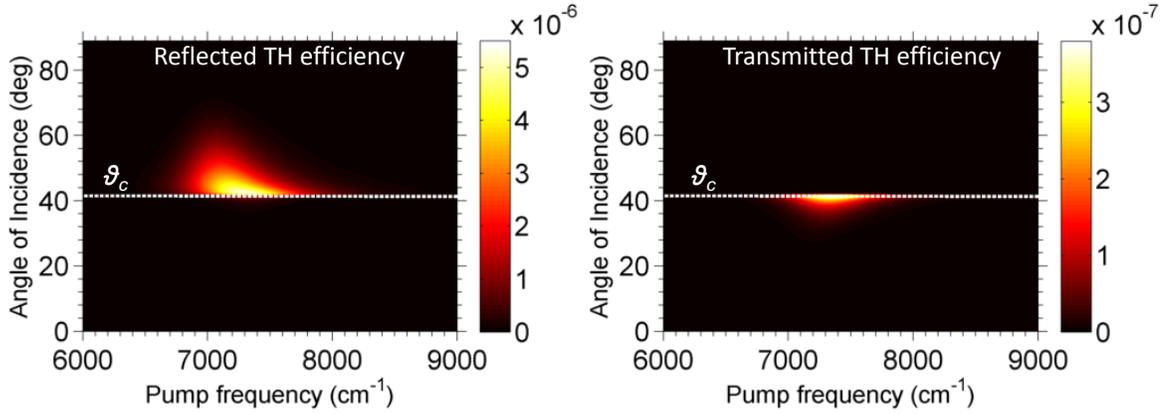

Fig. 4. (a) Reflected TH efficiency versus frequency and angle of incidence from full-wave simulations. (b) As in (a), for the transmitted TH efficiency. The dashed white lines indicate the critical angle of 41.8 degrees.

To further probe the origin of the enhanced third harmonic generation we compare the profile and magnitude of the electric field intensity within the ITO film for both the Kretschmann and non-Kretschmann excitation geometries. It is known that a thin plasmonic material supports long and short range surface plasmon modes[42]. As the thickness of the film shrinks into the deeply sub-wavelength regime, the long range surface plasmon mode evolves into the so-called ENZ mode, similar to the one discussed by Vassant et al.[13]. The ENZ mode utilized in the present work differs somewhat from that of Ref. 13, due to the difference in the substrate (metal in Ref. 13 and glass in the present work), but many of the important properties of the ENZ modes are similar for the two configurations. Figures 5a and 5b show the longitudinal electric field intensity (i.e. $|E_z|^2$) obtained from FDTD simulation[43] as a function of wavelength for an incidence angle of 43.7 degrees for the Kretschmann (Fig. 5a) and non-Kretschmann (Fig. 5b) excitation schemes (the transverse electric field intensity is negligible). Enhancement of the intensity within the film is clearly seen near the ENZ wavelength for both configurations. However, the field intensity is almost six times larger for the Kretschmann scheme, due to its ability to couple to the ENZ mode which lies to the right of the light line. Since the TH emission scales roughly as the intensity cubed, this observation is consistent with our experimental observation of ~200 times larger TH from the Kretschmann geometry. Further, we note that the field intensity is nearly constant across the ITO film, in agreement with what is expected for an ENZ mode[13].

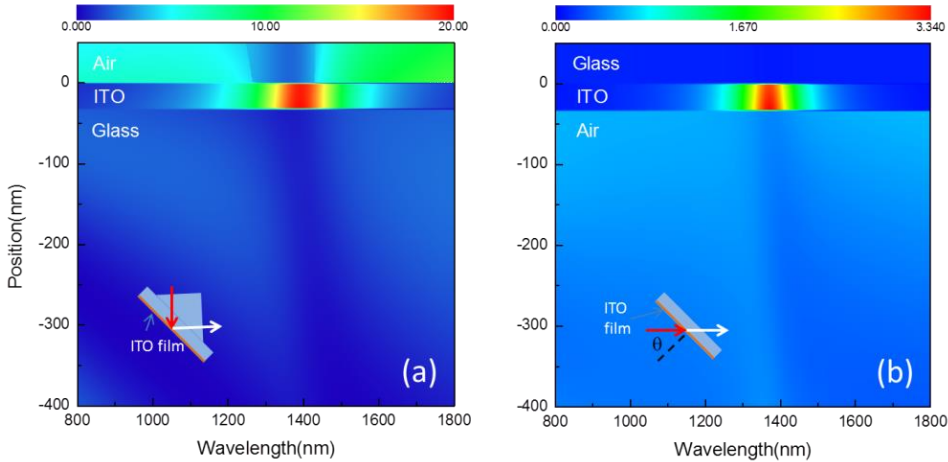

Fig. 5. (a) Spatial and wavelength dependence of the longitudinal intensity enhancement factor at an angle of incidence of 43.7 degrees for the Kretschmann geometry. (b) As in (a), but with the incident wave from the air side.

In conclusion, we have demonstrated that efficient third harmonic generation can be achieved with ultrathin ITO films by coupling the pump wave to the ENZ mode of the thin film. Exploiting the field enhancement effect resulting from efficient coupling to the ENZ mode, the TH yield is more than 200 times larger than when the pump wave is incident from air, and more than 10,000 times larger when only the glass prism is present. Because the ENZ wavelength is very sensitive to the electron density, the potential exists for active tuning via electrical modulation. Furthermore, the extremely small length scales involved render the need for phase matching irrelevant[33, 44]. Since the conditions used in this paper can be easily met in other plasmonic and low-loss reststrahlen materials, these results provide a general method for harmonic conversion for infrared and ultraviolet radiation in deeply sub-wavelength environments.

ACKNOWLEDGMENT
Portions of this work were supported by the U.S. Department of Energy, Office of Basic Energy Sciences, Division of Materials Sciences and Engineering, by the Laboratory Directed Research and Development program at Sandia National Laboratories, and were performed, in part, at the Center for Integrated Nanotechnologies, a U.S. Department of Energy, Office of Basic Energy Sciences user facility. Sandia National Laboratories is a multi-program laboratory managed and